\documentclass[epj]{svjour}
\usepackage{amsmath}
\usepackage{amssymb}
\usepackage{graphicx}
\usepackage{xcolor}
\usepackage{enumerate}
\usepackage{hyperref}
\def\cHI{\mathcal{H}_\text{I}}
\def\cHrb{\mathcal{H}_\text{I}^\text{(rb)}}
\def\cHrf{\mathcal{H}_\text{I}^\text{(rf)}}
\def\tc{T_\text{c}}
\def\Bc{B_\text{c}}

\newcommand{\argmin}{\mathop{\mathrm{argmin}}}

\begin{document}

\title{Temperature-dependent criticality in random 2D Ising models}
\author{Matteo Metra \and Luc Zorrilla \and Maurizio Zani \and Ezio Puppin \and Paolo Biscari}
\institute{Department of Physics, Politecnico di Milano, Piazza Leonardo da Vinci 32, 20133 Milan, Italy}
\date{\today}
\abstract{We consider 2D random Ising ferromagnetic models, where quenched disorder is represented either by random local magnetic fields (Random Field Ising Model) or by a random distribution of interaction couplings (Random Bond Ising Model). In both cases we first perform zero- and finite-temperature Monte-Carlo simulations to determine how the critical temperature depends on the disorder parameter. We then focus on the reversal transition triggered by an external field, and study the associated Barkhausen noise. Our main result is that the critical exponents characterizing the power-law associated with the Barkhausen noise exhibit a temperature dependence in line with existing experimental observations.
}

\maketitle

\section{Introduction}
\label{intro}

The bursty and intermittent character of natural phenomena is a fingerprint of complex, collective evolution, often characterized by the presence of quenched disorder, frustration, and a multiplicity of free-energy local minimizers. Quenched disorder requires that agents do not modify in time the rule they obey, but the parameters characterizing the individual rules vary from agent to agent. Frustration arises from the clash among individual goals. It prevents from expanding the choices which optimize local interactions to a global, unique ground state configuration. Finally, the presence of very many (infinite, in the thermodynamic limit) metastable states is a key ingredient to trigger intermittency. Systems trapped in a non-optimal equilibrium configuration have the opportunity of gathering large amount of energy, which induces bursty events when subsequently released. Examples are ubiquitous, from Universe fragmentation \cite{86bro,2020pup} to earthquakes \cite{54gr} and paper crumpling \cite{96hose,01sethna}, down to microevolution in plastic dislocation flow \cite{15bis,16bis,19bis}.

In this paper we focus on the reversal transition which occurs in ferromagnetic solids subject to a slowly varying external magnetic field. Heinrich Barkhausen \cite{19bark} first observed the intermittent character of the ferromagnetic response in such experiment. The intensity of the magnetization reversal bursts (named after him thence on) obeys a characteristic power-law distribution which extends over several orders of magnitude. Power-law distributions are in fact a fingerprint of intermittent phenomena \cite{96bak}. Their scale-free, heavy-tail character evidences the on-the-fly search for new metastable configurations in complex systems when external actions perturb the previous equilibria.

Zero-temperature simulations confirm that 2D Random Field Ising models evidence a critical behavior in the reversal transition either in both square \cite{99fron,11spas} and triangular \cite{17janicevic} lattices. The disorder-dependence of the critical temperature has been also reported in square ferromagnets under the effect of bimodal local fields ($\pm h_0$), chosen at random \cite{09cro}. It is our aim to confirm the above results by considering both a random-field and a random-bond square Ising model. By performing also finite-temperature simulations we are also able to study the temperature dependence of the critical exponent characterizing the power-law distribution of the Barkhausen noise.

We study two different random Ising models: the random-field (RFIM) and the random-bond (RBIM) Ising model. In both cases we consider a two-dimensional set of $N$ Ising spins $\{s_i=\pm1$, $i=1,\dots,N\}$, which occupy the vertices of a square lattice. Periodic boundary conditions are enforced to mitigate the finite-size effects. The spins evolve according to the Hamiltonian
\begin{equation}\label{ham}
\cHI[s]=\sum_{\langle i,j\rangle} J_{ij} s_i s_j + \sum_{i=1}^N B_i s_i,
\end{equation}
where the first sum is performed over pairs of neighboring spins.

In the RFIM all the coupling constants $J_{ij}$ are set equal to a common value $J_{ij}=J$, which is chosen to be positive to model a ferromagnetic system, while the magnetic fields $B_i$ are random. Their values are extracted from a Gaussian probability distribution whose mean value $B$ is a control parameter (mimicking the external magnetic field), while the standard deviation $\sigma$ models the presence of possible defects and quenched fields in the lattice, and is assumed to be independent of the imposed temperature and external magnetic field. In the RBIM the randomness is assumed to affect the spin interactions instead. As a result, the coupling constants $J_{ij}$ are extracted from a Gaussian distribution of mean value $J>0$ and standard deviation $\sigma$, while the magnetic field is uniform: $B_i=B$. In the RBIM some spin pairs might be antiferromagnetic, depending on the relative value of $\sigma$ and $J$. The ferromagnetic ground state remains stable against transition to a frustrated (spin-glass) phase as long as the probability of negative coupling constants is low enough ($p<0.15$, see \cite{80ja}). In all our simulations we will be on the ferromagnetic side of this limit. In both cases, we call \emph{disorder} parameter the standard deviation $\sigma$, and label as $\cHrf$ and $\cHrb$ the Hamiltonians corresponding to the RFIM and RBIM choices.

In our analysis we report and analyze the results of two specific numerical experiments. The first is the temperature-driven phase transition between the ferro- and the paramagnetic phases. The aim of this analysis is to understand how randomness affects the Curie temperature $\tc$. In the second experiment we simulate the transition between two opposite ferromagnetic ground states, mediated by a reversal of the external magnetic field. We will specifically focus on the intermittent character of this transition, in order to characterize the statistical properties of the associated Barkhausen noise.

The intermittent character of the spin-reversal transition was first experimentally measured in bulk 3D polycrystals \cite{00durzap} and thin films \cite{00puppin}. The presence of Barkhausen noise has been related to the dynamics of ferromagnetic domain walls in the depinning transition \cite{98zapdur}. Further experimental studies evidenced a number of properties that will be investigated in the present study. These include the following.
\begin{itemize}
\item Particular attention is focused on the critical exponent characterizing the Barkhausen noise distribution. Such coefficient $\alpha$ is defined by the power law distribution for the amplitude $\Delta m$ of the magnetization jumps taking place inside the material during the magnetization process: $P(\Delta m) \sim (\Delta m)^{-\alpha}$. The exponent $\alpha$ exhibits a remarkable temperature dependence \cite{04pupzan,04zanpup}. More precisely it raises from $\alpha=1$ at room temperature to $\alpha=1.8$ at low temperatures ($T=10\,$K).
\item The critical exponents do not vary significantly with the sample width \cite{07puppino}. This suggests that the effects are intrisically two-dimensional, and supports the choice of conducting 2D simulations to inspect how the material properties influence the intermittent response to the variations of the external field.
\item The \emph{coercive field}, defined as the value of the external field at which the overall magnetization changes its sign, exhibits a temperature dependence \cite{04pupzan} which agrees with a theoretical prediction obtained by assuming that the transition is governed by domain-wall motion \cite{83gaunt}.
\end{itemize}
The present study aims at analyzing if and how randomness, quenched either in the interactions or in the magnetic response, succeeds in reproducing the measured behavior. In order to easy notations, in the following we will set the average coupling constant $J=1$ and the Boltzamnn constant $k_\emph{B}=1$. This amounts to set $J$ as unit measure for energies, magnetic fields and disorder parameter, and to measure the temperature $T$ in units of $J/k_\emph{B}$.

The plan of the paper is as follows. In the next section we study how the ferro-para (Curie) transition temperature is influenced by the disorder parameter $\sigma$. In Section~\ref{sec:Bark} we focus on the Barkhausen noise that characterizes the reversal transition. In the concluding section we review and discuss the main results here reported.

\section{Curie Temperature in Random Ising Models}
\label{sec:curie}

\subsection{Classical Ising system}

We start by briefly reporting the results of the tests performed for a simple 2D Ising model, in the absence of any disorder ($\sigma=0$). This proves to be useful in understanding which quantities might be of help in the following to better detect the Curie temperature $\tc$, that is the temperature at which the para-ferromagnetic transition occurs.

For a 2D square Ising ferromagnet, Onsager \cite{44onsa} computed the exact temperature dependence of the average spontaneous magnetization $m$ (in the thermodynamic limit), as well as the analytic value of $\tc$ itself
\begin{equation}\label{mons}
m(T)=1-\sinh^{-4}\frac{2}{T} \quad\text{for }T\leq \tc,\quad\text{with}\quad \tc=\frac{2}{\log(1+\sqrt{2})}=2.269...
\end{equation}

In finite-size systems, however, computing the average magnetization might not be the most efficient way to detect the value of the Curie temperature. In a pure finite size non-disordered Ising system, a better estimate may be performed by studying the system's response to a slowly varying external field. A classical Ising system behaves in a quite predictable way in such experiment. In the ferromagnetic phase there are two equilibrium states. Whenever the external magnetic field is reversed, the equilibrium state which was the ground state becomes metastable, and viceversa.

If the chosen dynamics involves one or only a few spins at a time, the system might remain trapped in a metastable state for an unlimited time. For example, at null or very low temperatures, the external magnetic field must overcome the local field induced by the four neighboring spins $(B_\text{loc}=4J)$ to trigger the transition. Let us label as \emph{coercive} the external field $\Bc(T)$ needed to reverse the average magnetization starting from a perfectly aligned configuration, so that $m(B,T)\lessgtr0$ when $B\lessgtr\Bc(T)$. Then, $\Bc(0)=B_\text{loc}$ if the system evolution explores only the immediate surroundings of the present configuration. A positive value of the coercive field generates an hysteresis loop when the external field is first increased until a complete positive magnetization is obtained, and then decreased back to the initial state. We remark that the coercive field would vanish if we allowed the system to explore all the configuration state, as in this case metastable states would be immediately abandoned in favor of the state with a magnetization coherent with the external field.

\begin{figure}
\begin{center}
\includegraphics[height=6cm]{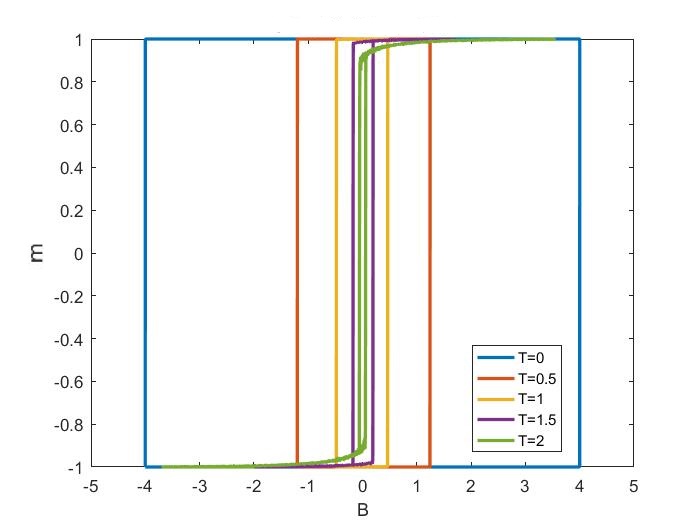} \qquad
\includegraphics[height=6cm]{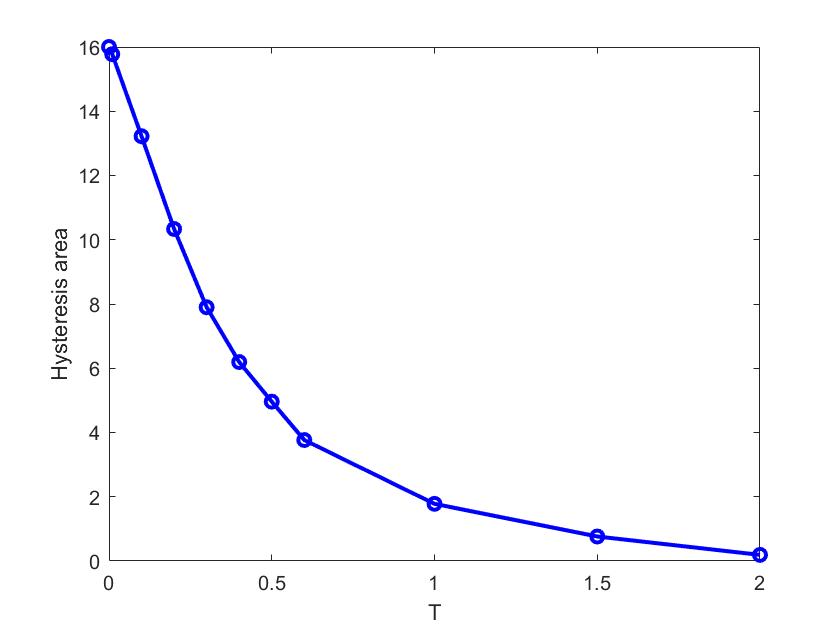}
\caption{[Left] Hysteresis loops of the spontaneous magnetization as a function of a varying external magnetic field. Different colors represent different temperatures, as evidenced in the inbox.  \hfill\break
[Right] Hysteresis area $A$ enclosed in the hysteresis loops observed by increasing and then decreasing the external field, as a function of the temperature. The area is measured in units of $J$.}\label{fig:hyst}
\end{center}
\end{figure}

Figure~\ref{fig:hyst} displays some hysteresis loops, and quantifies the area $A$ enclosed by them, for several different temperatures.
All the simulations here reported have been conducted by using a single-flip Metropolis algorithm run on a 2D square Ising system composed by $10^4=100\times 100$ spins. Thermalization was ensured by collecting data after $10^8$ Monte-Carlo steps, which in average corresponds to $10^4$ Monte-Carlo steps per spin. Further details on the numerical simulations are reported in \cite{tesimetra}.

The analysis in Figure~\ref{fig:hyst} refers to a pure Ising system, in which disorder is still to be introduced. As commented above, the choice of a single-flip Metropolis algorithm generates a non-null coercive field. The plots evidence that its value decreases with the temperature, as the thermal fluctuations allow the system to explore the phase space most efficiently and the new ground state is more easily found. In any case, at all ferromagnetic temperatures the reversal transition remains abrupt, with $\Bc>0$. At the Curie temperature and above, $\Bc$ vanishes, the transition is smoothed, and the hysteresis loops disappear. The qualitative feature of the reversal transition (abrupt vs.\ smooth) proves to be an efficient tool to identify the critical temperature in the presence of disorder.

\subsection{Random-Field and Random-Bond Ising Model}

The random Ising models have been since long presented as archetypal examples of systems where the reversal transition described above is expected to occur through a sequence of magnetization jumps originated by spin avalanches \cite{93seth}. Before entering the statistical analysis of the magnetization jumps, we will here focus on the effect of the disorder parameter on the para-ferro temperature transition. The results reported in in the following of the present paper confirm that the Curie temperature $\tc$ decreases with $\sigma$, a result which is in line with mean-field \cite{82vanH} and 2D calculations \cite{96cho}.
From the physical point of view it implies that in the presence of quenched disorder the system needs less thermal energy to destroy the long-range
ferromagnetic order.

In a disorderd Ising system the method adopted previsously in the absence of disorder (based on the detection of a discontinuity in the magnetization curve) is not any more adequate. Instead, it is more suitable  to follow the divergence of the magnetic susceptibility $\chi=\partial m /\partial B$ at the reversal transition. The left panel of Figure~\ref{fig:hyst} shows indeed that the reversal transition is basically immediate in all the ferromagnetic phase. Nevertheless, the quenched disorder introduces intermittency in the reversal transition, and the detailed shape of the magnetization \emph{vs.} external field curves vary significantly from test to test. We need to introduce a robust procedure to extract detailed information from these curves (such as the partial derivative at the reversal point involved in the magnetic susceptibility).

\begin{figure}
\begin{center}
\includegraphics[height=6cm]{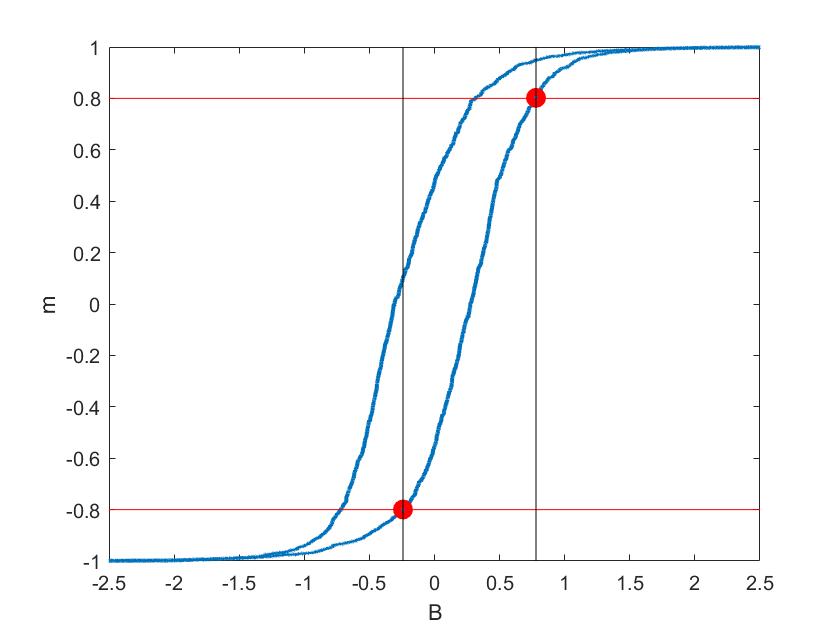} \qquad
\includegraphics[height=6cm]{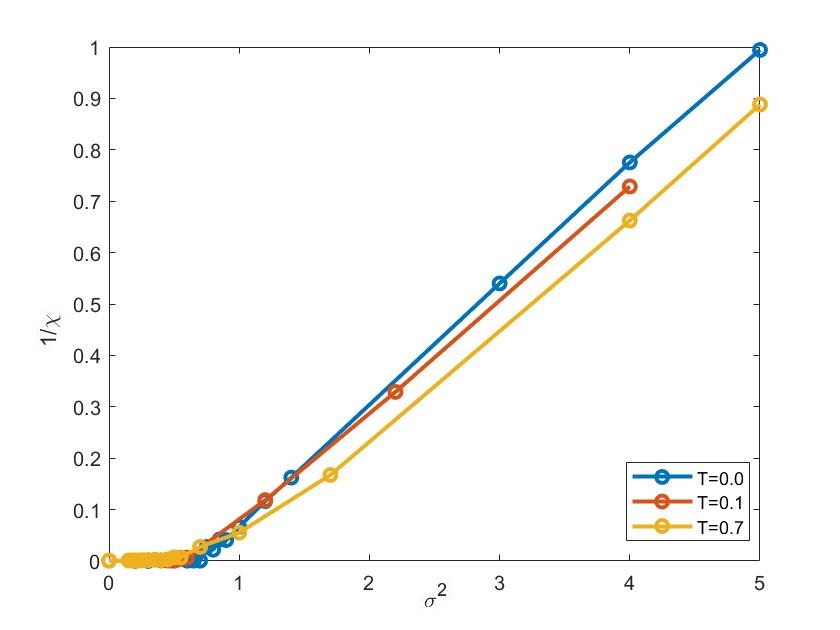}
\caption{[Left] Magnetization and external field values used to estimate the inverse susceptibility during the reversal transition. The simulation reports the magnetization for the RFIM with $\sigma=2$ and $T=0.7$\hfill\break
[Right] Inverse susceptibility as a function of the disorder parameter for different temperatures. The inverse susceptibility vanishes in the ferromagnetic phase.}\label{fig:susc}
\end{center}
\end{figure}

We introduce a discrete estimate of the magnetic susceptibility, computed as illustrated by the left panel of Figure~\ref{fig:susc}. The figure shows a typical reversal graph at a temperature above-critical. Notice that in the presence of disorder a non-vanishing hysteresis loop can be spotted also in the paramagnetic phase, as the random fields (or bonds) anticipate the start of the reversal transition with respect to an ordered system. In order to estimate the zero-field magnetic susceptibility, we arbitrarily choose two fixed magnetization values (resp.\ $m=\pm0.8$, with $\Delta m=1.6$), and compute in all cases the difference $\Delta B=B_2-B_1$ between values of the external field at which the chosen magnetization values are crossed for the first time. The resulting inverse magnetic susceptibility, defined as
\begin{equation}\label{eq:invsusc}
\chi^{-1}\approx\frac{\Delta B}{\Delta m}
\end{equation}
is a much more stable quantity, and can be averaged over a limited number of tests (10 in the points displayed in the right panel of Figure~\ref{fig:susc}) to obtain a clear trend of how the susceptibility depends on $\sigma$ and $T$. In particular, this panel evidences that the inverse susceptibility has a linear dependence on the square of the disorder parameter $\sigma$ in the paramagnetic phase, and vanishes at the transition. A linear fit of the paramagnetic data allows then to locate quite precisely the para-ferro transition.

\begin{figure}
\begin{center}
\includegraphics[height=6cm]{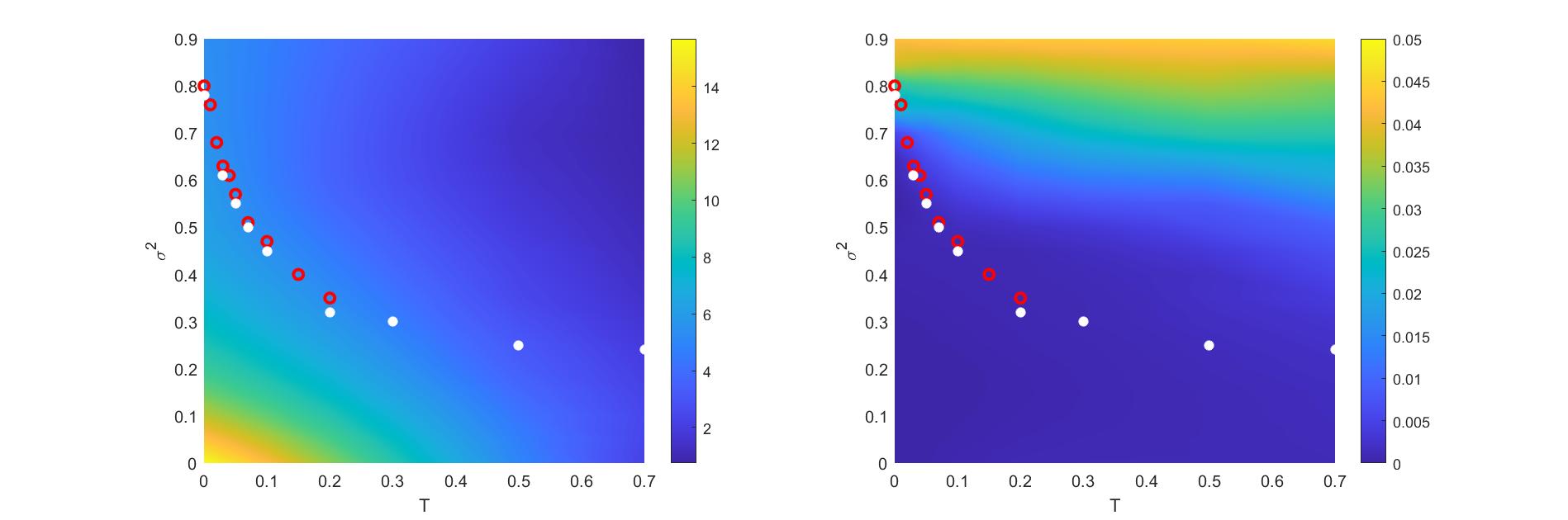}
\caption{Color plots of the hysteresis area (left) and the inverse susceptibility (right) for different temperatures and disorder parameters for the RFIM. The white and red dots in both the plots provide the better estimates of the temperature-dependent critical disorder obtained through other methods, described in the text. More precisely, the white dots originate from the best fit of the inverse susceptibility curves, while the red dots come from the statistical analysis of the Barkhausen jumps discussed in the next section.}\label{fig:pcolor}
\end{center}
\end{figure}

Figure~\ref{fig:pcolor} displays two color plots reporting the RFIM hysteresis area $A(T,\sigma)$ (left) and the slope of the paramagnetic dependence of the inverse susceptibility as in (\ref{eq:invsusc}) on the temperature and the disorder parameter. The white dots in the plots display the critical temperature and disorder values for the para-ferro transition as computed from a linear fit of the inverse susceptibility curves. The curves display also some red dots, which correspond to the best estimate of the critical temperature that can be obtained by analysis the statistical properties of the Barkhausen noise as discussed in the next section. All the data draw a coherent picture from which it emerges that the critical temperature decreases when the disorder parameter increases.
In particular, the white squares and red dots identify the para-ferro transition for the RFIM.
Qualitatively similar plots (not reported here for brevity) can be obtained with the RBIM.

\section{Barkhausen analysis}
\label{sec:Bark}

The main focus of the present work is to check whether the random Ising models mimic properly the intermittent statistical properties measured during the reversal transition.

\subsection{Zero-temperature results}

The $T=0$ single-flip Monte-Carlo simulations are deterministic. At each step, a move is accepted if and only if it generates an energy gain ($\Delta\cHI\leq 0$). We consider a simulation which starts from the fully oriented configuration corresponding to extremely large and negative external magnetic field (\{$s_i=-1$ for all $i$\}), and we aim at tracing the response when $B$ is slowly increased.
In the zero-temperature regime it is possible to predict exactly the value of the external field at which every spin flips. Since the only allowed transitions are those which lower the system energy, the spins will be allowed only to flip from their negative original value as the external field increases. Opposite transitions $(+1 \to -1)$ are all hindered in this regime. Let us rewrite the local magnetic field and the interaction couplings in order to separate their average value from the disordered contribution
\begin{equation}\label{eq:bandj}
B_i=B+\widetilde{B}_i \quad \text{(RFIM)}, \qquad J_{ij}=J+\widetilde{J}_{ij} \quad \text{(RBIM)},
\end{equation}
where $\widetilde{B}_i$ and $\widetilde{J}_{ij}$ are now a set of zero-mean Gaussian parameters. For both the RFIM and the RBIM, the energy variation associated with a $-1 \to +1$ flip of the $i$-th spin is given by
\begin{equation}\label{eq:imin}
\Delta \mathcal{H}_{\text{I},i}=-2\left(B+\widetilde{B}_{i} +\sum_{<i,j>}(J+\widetilde{J}_{i j})s_j\right),
\end{equation}
where as usual the sum in the right-hand term is performed over the spins $j$ which are nearest neighbors of the chosen spin $i$.
Since we start from a negative-oriented configuration, the first spin (say, the $i$-th) would flip if
\begin{equation}\label{eq:imin2}
B\geq B_{\text{cr},i}=4J-\widetilde{B}_{i}+\sum_{<i,j>}\widetilde{J}_{i j},
\end{equation}
where we have considered that the spin $i$ has four nearest neighbors.
In particular the first spin to flip, say the $\bar{\text{\i}}$-th, is reversed when
\begin{equation}\label{eq:imin3}
B=B_\text{cr}=\min_i B_{\text{cr},i}, \quad \text{with}\quad
\bar{\text{\i}}=\argmin\limits_iB_{\text{cr},i} .
\end{equation}
Every time a spin flips, the critical external field values of its four neighbors $j$ vary. More precisely, from \eqref{eq:imin} it follows that when $s_{\bar{\text{\i}}}$ flips from -1 to +1, $B_{\text{cr},j\text{\,(old)}}$ is modified into
\begin{equation}\label{eq:imin4}
B_{\text{cr},j\text{\,(new)}}=B_{\text{cr},j\text{\,(old)}}-2J_{\bar{\text{\i}} j}
\end{equation}
In particular, the critical fields are lowered for all those neighboring spins which were connected to the flipped one through ferromagnetic links. If $B_{\text{cr},j\text{\,(new)}}\leq B_{\text{cr},\bar \i}$ the spin $j$ is dragged by the spin $\bar{\text{\i}}$ and flips at the same value of the external field. This possibly originates a spin avalanche, as several spins might flip at the same value of the external field.

In the following we examine quantitatively two features of this bursty process by performing two different simulations.
\begin{enumerate}
\item In the first (that we discuss below for the RFIM) we vary the external field at fixed intensity steps $\Delta B$, and follow the ensuing magnetization jumps. We remark that within the same external field variation $\Delta B$ several avalanches may be gathered, so we expect to find magnetization jumps larger than those evidenced in the single-avalanche studies.
\item In the second we monitor the intensity of the individual avalanches occurring when the external field achieves a critical value, follow the evolution of the single avalanches, and then study also the distance between consecutive critical external field values. This latter study allows to inspect the \emph{interevent} distribution, and to compare it to similar distributions that have been predicted or measured for recurrence times in earthquakes \cite{09talb}, human communication \cite{15panza}, or theoretical models \cite{17jo}.
\end{enumerate}

\begin{figure}
\begin{center}
\includegraphics[height=6cm]{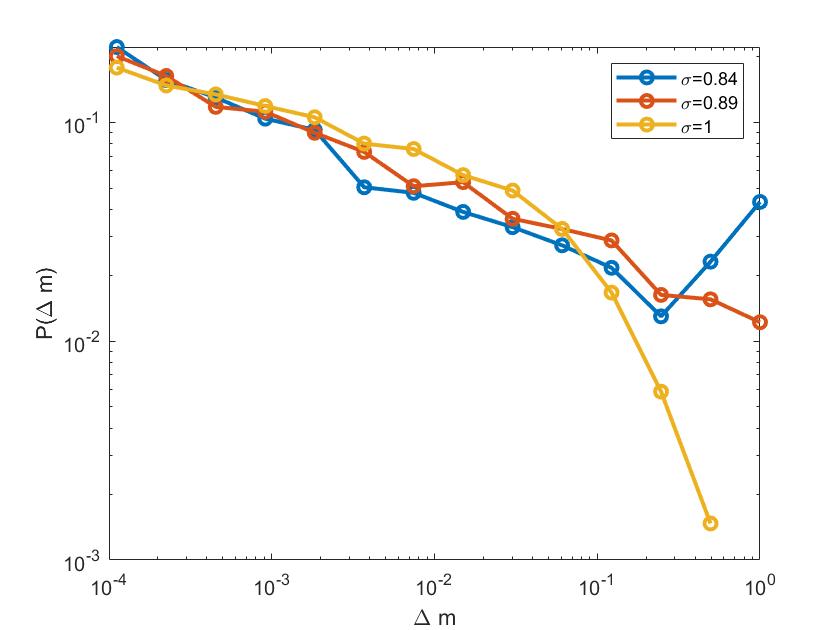}
\caption{Magnetization jumps probability distribution for three different values of the disorder parameter $\sigma$ in the RFIM: $\sigma=1.00$ (yellow line), 0.89 (red), and 0.84 (blue). The yellow and blue distributions have been shifted vertically to evidence how the left part of the distribution shares the same properties at all values of $\sigma$.}\label{fig:barkt0}
\end{center}
\end{figure}

Figure~\ref{fig:barkt0} displays three different log-log histograms reporting the probability of having the specified magnetization jumps in a RFIM in a type-(i) experiment, that is when the external field is varied by a fixed step ($\Delta B=10^{-3}$). The effect of the disorder parameter $\sigma$ on the jump distribution emerges clearly. When the disorder is too low (blue line) the collective behavior prevails, and very large jumps become increasingly probable. On the other hand, when the disorder to too high (yellow line) the collective response is lost. The spins tend to evolve indepedently, and large jumps become rare. The two regimes are separated by a special, critical value of the disorder parameter, at which the complex power-law behavior prevails at all jump intensities. The power-law characteristic of Barkhausen noise emerges clearly from the data, with a critical exponent slightly larger than $0.3$.

\begin{figure}
\begin{center}
\includegraphics[height=5.5cm]{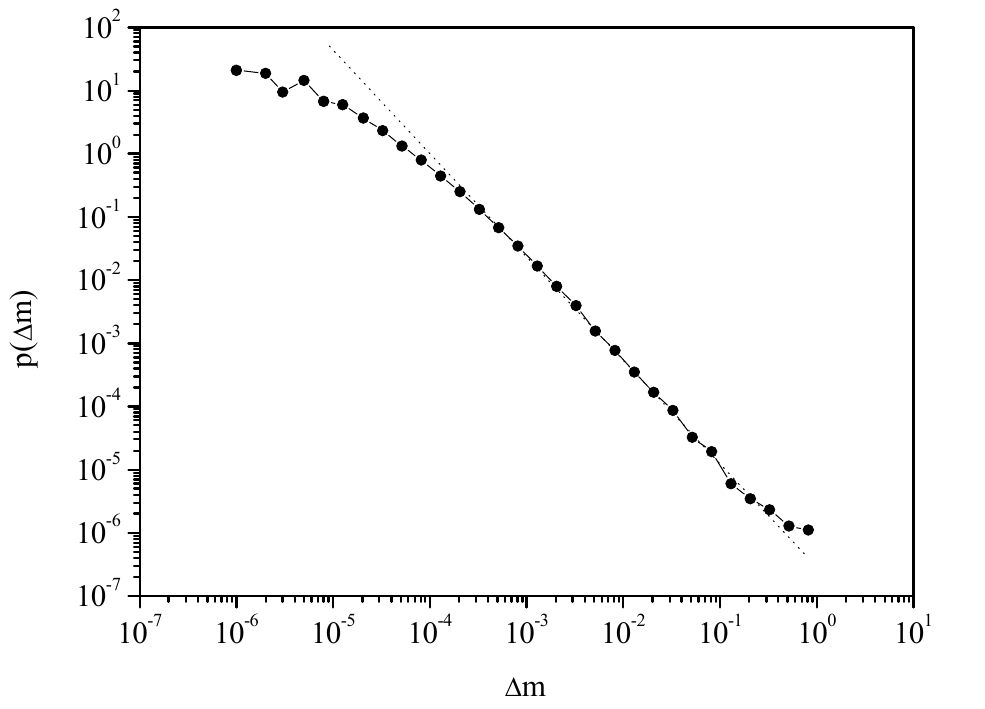}
\caption{Magnetization jump distribution at $T=0$ in the RFIM obtained by counting separately each avalanche (and not gathering any avalanches in a larger event), independently of the distance between the external field values at which they occur. The best fit for the critical exponent is slightly smaller than 2, fairly in accordance with experimental measures.}
\label{fig:Zani}
\end{center}
\end{figure}

The value estimated for the critical exponent in the reversal transition depends on how single jumps and avalanches might be gathered into a single, possibly significantly larger, abrupt event. Indeed, if we were able to follow individual avalanches several apparently large avalanches would split into a bunch of smaller jumps. As a result, in the type-(ii) experiment described above large events are expected to become less probable, and a larger critical exponent should be registered. Figure~\ref{fig:Zani} confirms our expectations. In it, the probability of occurrence of individual avalanches is reported for a large (1000 x 1000 spins) RFIM at $T=0$ and the critical disorder. The critical exponent turns out to be slightly smaller than 2, and therefore much larger than the one reported in type-(i) experiments. The fact that this computed value is much closer to the experimental measures denotes that these latter are indeed performed at \emph{quasi}-equilibrium, with the external magnetic field varying so slowly that the experiment records a series of basically individual events.

\begin{figure}
\begin{center}
\includegraphics[height=4cm]{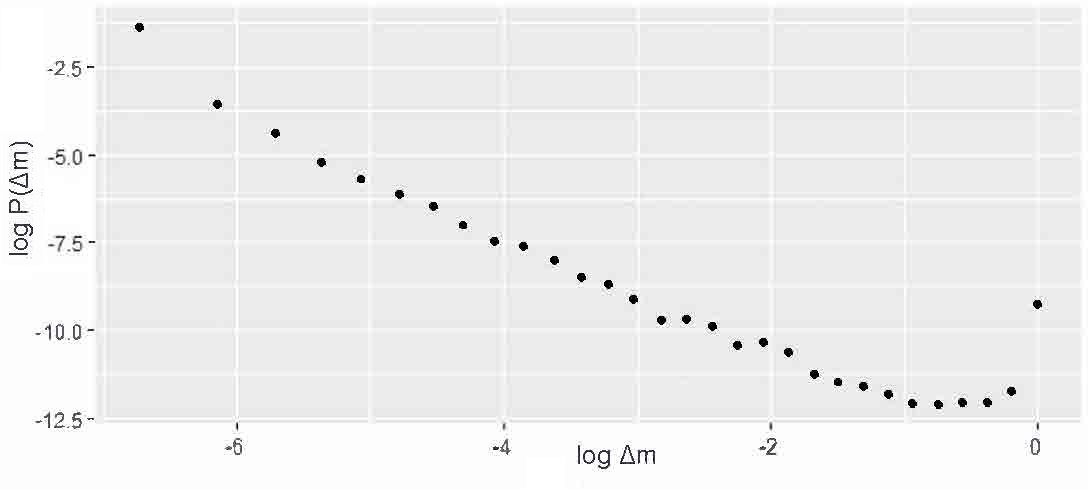} \qquad
\includegraphics[height=4cm]{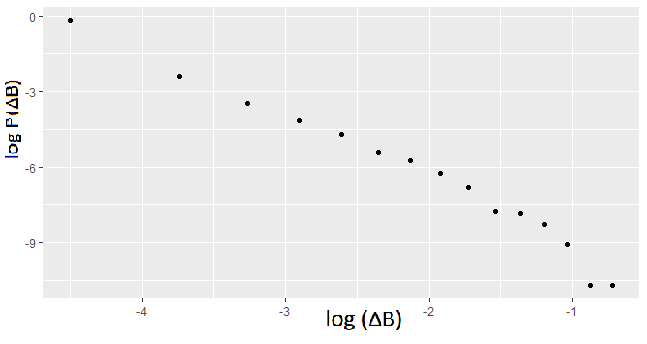}
\caption{[Left]: Magnetization jumps during the reversal transition in a RBIM at $T=0$ with $\sigma=0.48$. \hfill\break
[Right]: Probability distribution of the external field difference between two distinct magnetization avalanches in a RBIM at $T=0$.}
\label{fig:Zjumps}
\end{center}
\end{figure}

Figure~\ref{fig:Zjumps} reports the results of a similar type-(ii) experiment on a RBIM at a slightly above-critical value of the disorder parameter ($\sigma=0.48$). The distributions refer thus to single-avalanche monitoring. The left panel confirms that the Barkhausen-like power law distribution emerges also in the single-avalanche experiment for a RBIM for more than four decades in magnetization jumps intensity. The qualitative features of the distribution in the plot coincides with the type-(i) experiment shown in Figure~\ref{fig:barkt0}. Again, the value of the critical exponent varies significantly between the single-avalanche simulation in Figure~\ref{fig:Zjumps} (slope larger than 1.5) and the multiple-avalanches simulation in Figure~\ref{fig:barkt0} (slope smaller than 0.5).
The right panel of Figure~\ref{fig:Zjumps} shows also the \emph{interevent} distribution, that is the probability of finding a specific $\Delta B$ between any two single avalanches in a type-(ii) experiment. It also follows a power-law distribution as observed in earthquakes, human communication, and other complex systems \cite{09talb,15panza,17jo}. The fact that the intervent distribution obeys a power-law suggests that very small $\Delta B$'s should be chosen if we intended to avoid the gathering of small avalanches into larger events.

\subsection{Finite temperatures}

At low, yet finite temperatures the single-flip Monte-Carlo algorithm we used still provides reliable results. Nevertheless, as the temperature is increased, the single-flip algorithm struggles in its effort to describe properly the statistical features of the reversal transition. We postpone to the final section the discussion of the motivations behind this lack of precision, along with the possible remedies that could improve the results up to, or above the critical temperature.

\begin{figure}
\begin{center}
\includegraphics[height=5.5cm]{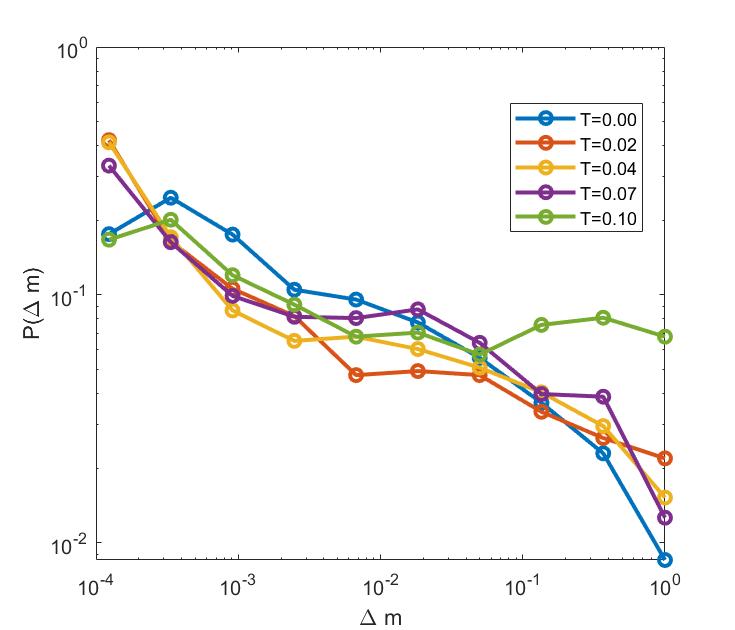} \qquad
\includegraphics[height=5.5cm]{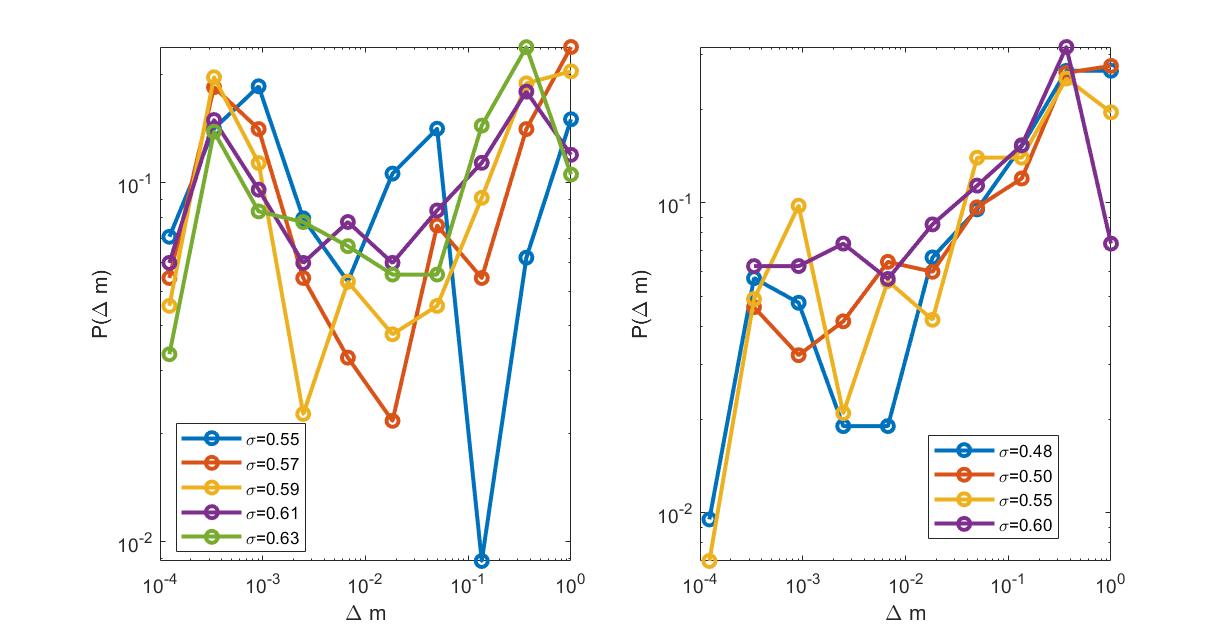}
\caption{Magnetization jump distributions obtained for different values of the temperature and disorder parameter in the RFIM. \hfill\break
[Left]: Low-temperature distributions ($T$ values shown in the panel) at the critical value of the disorder parameter, that is the value of $\sigma$ at which the power-law behavior spans most decades. \hfill\break
[Middle]: Jump distributions at $T=0.2$, for different values of the disorder parameter. The critical value for $\sigma$ is close to $0.60$ (see also Figure~\ref{fig:pcolor}) \hfill\break
[Right]: Jump distributions at $T=0.4$, for different values of the disorder parameter. The critical value for $\sigma$ is about $0.55$ (see also Figure~\ref{fig:pcolor}).}
\label{fig:finiteT}
\end{center}
\end{figure}

Figure~\ref{fig:finiteT} displays the result of the magnetization jump analysis performed at different temperatures for the RFIM. For the sake of briefness we do not report the (very similar) results that can be obtained for the RBIM. The left panel collects the distributions obtained at the critical values of the disorder parameter, from zero-temperature up to $T = 0.1 $. Notice that these must still be considered very low temperatures, as the Curie temperature in the absence of disorder is greater than 2.26 in the same units. The magnetization jump distributions at these temperatures exhibit a critical-disorder behavior that may still be labelled as a power-law over almost four decades. The (green) distribution corresponding to the largest temperature in the plot already announces that by increasing the temperature the probability of large jumps increases significantly. This trend is fully confirmed by the distributions shown in the middle and right panels, corresponding respectively to $T=0.2$ and $T=0.4$. Besides the emergence of an increasing noise that makes it difficult to extract properly averaged distributions, the most remarkable effect is that large jumps become more and more frequent until they dominate the distribution. This odd behavior, which is related to the difficulty of the single-flip algorihm to trace individual avalanches, will be discussed in the final section.

\subsection{Critical exponent}

\begin{figure}
\begin{center}
\includegraphics[height=5.5cm]{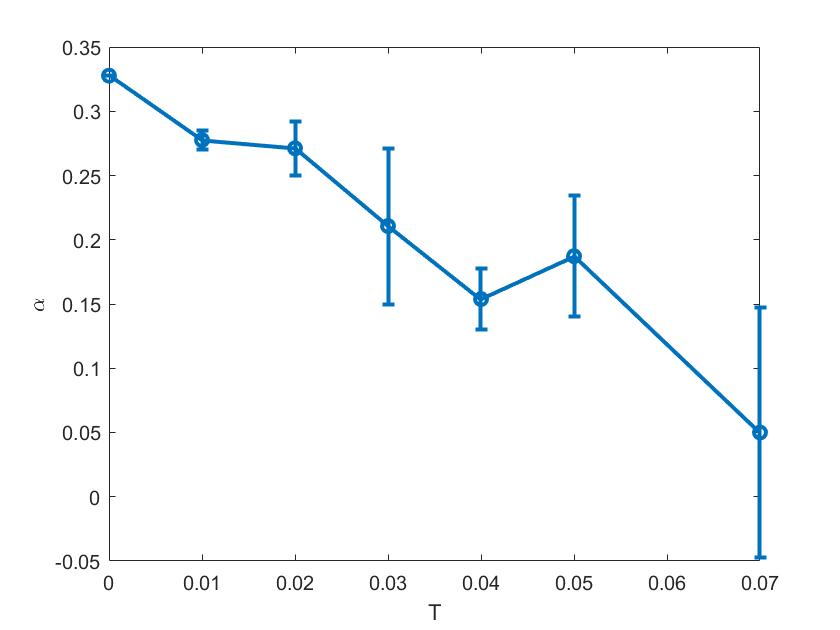} \qquad
\includegraphics[height=5.5cm]{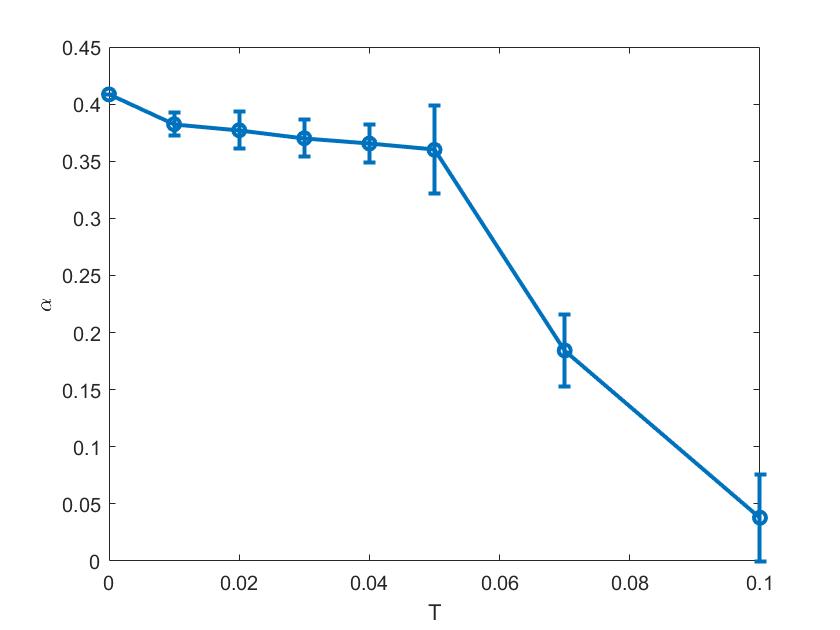}
\caption{Temperature dependence of the critical exponent $\alpha$ characterizing the magnetization-jump power-law distribution in the RFIM (left panel) and in the RBIM (right). The remarkable increase in the fitting error is related to the lack of power-law shape of the related distributions (see also Figure~\ref{fig:finiteT}).}
\label{fig:coeff}
\end{center}
\end{figure}

Experimental data indicate that the critical exponent $\alpha$ characterizing the Barkhausen noise distribution decreases with the temperature \cite{04pupzan,04zanpup}. Figure~\ref{fig:coeff} shows how the critical exponent depends on the temperature in the temperature range examined above. Both the RFIM (left) and the RBIM (right) data suggest that $\alpha$ decreases with the temperature in the type-(i) experiments described above. Though this trend is in accordance with the experimental observations, the value of the critical exponent is significantly different (it ranges from 0 to 1, not in accordance with the experimental measures, which range from 1 to 2). This discrepancy suggests that large events are overcounted by the algorithm we adopted. Indeed, every time we induce a magnetic field variation $\Delta B$ a new equilibrium configuration is found by looking from possible reversals in all the sample regions. This generates multiple avalanches, which are gathered in the same count, and therefore increase the probability of larger magnetization jumps. This interpretation is confirmed by the distributions displayed in left panel of Figure~\ref{fig:Zjumps} (RBIM) and in Figure~\ref{fig:Zani} (RFIM), where the jump distribution is reported at zero temperature RBIM by counting separately each avalanche in a type-(ii) experiment. In both distributions the critical exponent exceeds 1.5, and approaches the experimentally measured values.

\subsection{Coercive magnetic field}

\begin{figure}
\begin{center}
\includegraphics[height=5.5cm]{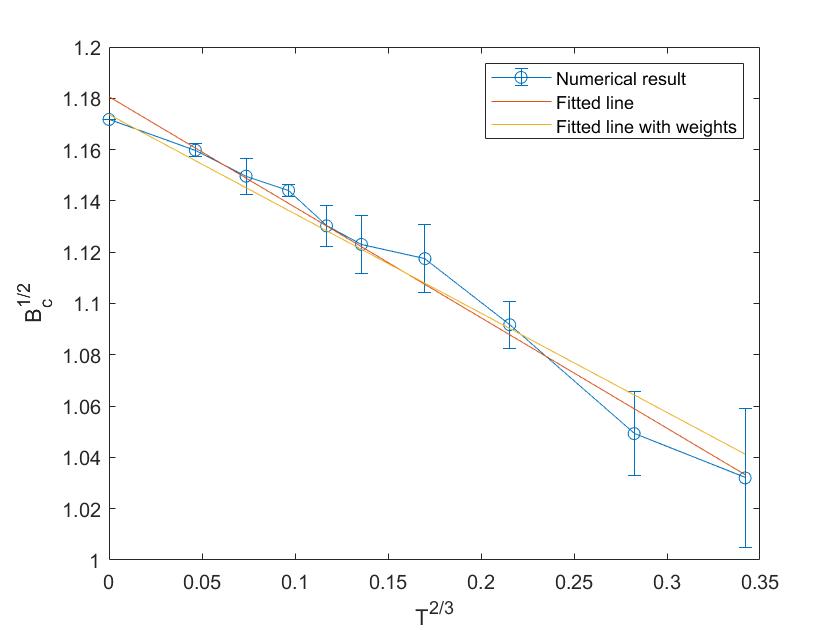} \qquad
\includegraphics[height=5.5cm]{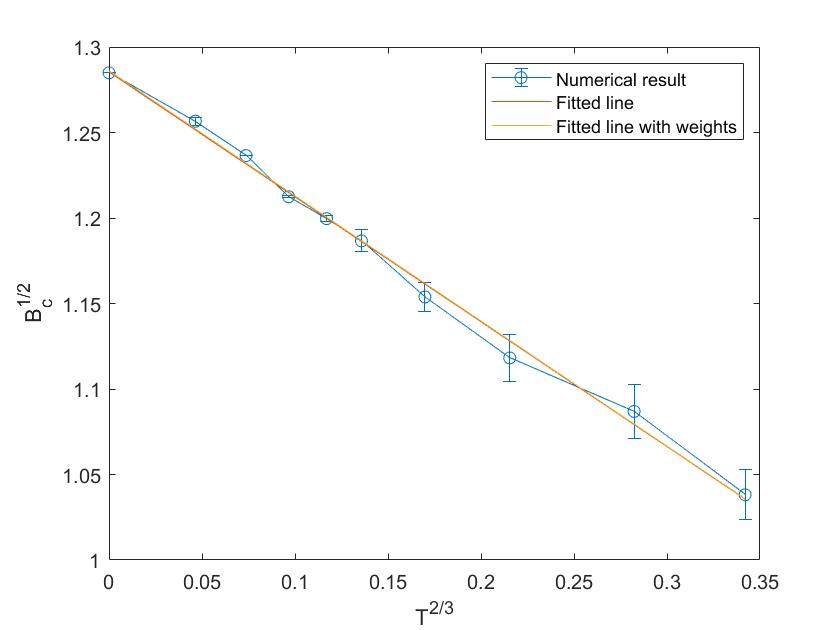}
\caption{Temperature dependence of the coercive field (defined in the text) both the RFIM (left panel) and the RBIM (right).}
\label{fig:coerc}
\end{center}
\end{figure}

The \emph{coercive field} is defined as the value of the external field at which the overall magnetization changes its sign. Figure~\ref{fig:coerc} evidences that the results obtained in both the RFIM and the RBIM agree with the theoretical prediction in \cite{83gaunt}: in both cases the square-root of the coercive field depends linearly from $T^{2/3}$. Notice that the theoretical calculations were performed by assuming that the transition is governed by domain-wall motion, an hypothesis absent in the present study, where the Monte-Carlo flips are in principle allowed to emerge anywhere in the material.

\section{Discussion}

We have analyzed the magnetization reversal transition in random Ising models,and its temperature dependence. Our results confirm that quenched randomness is a plausible origin for the intermittent Barkhausen noise which emerges in the reversal process. The distributions of magnetization jumps follow power law distributions which extend over several decades for a wide range of temperatures and disorder parameters.
The critical exponent associated with these distributions exhibits a temperature dependence in line with the experiments measurement. The temperature dependence of the coercive field is in agreement with earlier theoretical predictions.

Our simulations show that the evolution of random Ising models across (possibly metastable) equilibrium configuration captures the basic physical features of the intermittent reversal transition. It also emerges that at finite temperature the single-flip Monte-Carlo algorithm is not optimal in quantifying power-laws and related critical exponents (see Figure~\ref{fig:finiteT}). The main motivation for this is that the spins tested by a single-spin Monte-Carlo are chosen at random. Once a spin flips, the spins that are next inquired for possible further transitions are not necessarily (in fact, almost never) neighbors of the spin which was first reversed. This, by construction, hinders the formation of avalanches and is certainly not the optimal tool to monitor the domain reversal phenomena typical of the Barkhausen noise. True, the Monte-Carlo algorithm eventually reaches an equilibrium configuration. However, in view of the spin-choice issue above it is necessary to wait for so long until equilibrium is reached that several avalanches gather in the same counting. As a result, large magnetization jumps are favored, which explains the quantitative difference between the estimated values for the critical exponents (lower than 0.5) and the experimentally measured values (larger than 1).

In general, the Monte-Carlo algorithms ensure that an equilibrium configuration is eventually reached once the system is given enough simulation time. The problem is that a random ferromagnet possesses very many \emph{metastable} equilibrium configurations, but only two unique ground states (corresponding to a positive or a negative magnetization). In the presence of an external field, the ground state degeneracy is broken, with the preferred equilibrium magnetization agreeing with the external field. The intermittent response captured by the Barkhausen noise is triggered precisely by the wandering of the system across metastable states, until the ground state is eventually reached. This path is non-unique. In particular, if the system is allowed to evolve for a sufficiently long time, it is most probable that the equilibrium configuration attained coincides or is at least closer to the (unique) ground state, thus again favoring large jumps in the reversal process.

This drawback would be certainly reduced if not settled if we replaced the single-flip Monte-Carlo simulations with other (Cluster-like) Monte-Carlo algorithms. Ongoing simulations \cite{21perani} provide good confidence that more precise estimates for the critical exponents can be obtained up to the critical temperature. This is confirmed also by the zero-temperature results we have shown in Figures~\ref{fig:barkt0} and \ref{fig:Zjumps}. In both cases, we were allowed to monitor individual avalanches one by one, and the resulting critical exponent was more in line with the measured one.

To conclude, it is to be remarked that both RBIM and RFIM succeed in reproducing the bursty behavior of the reversal transition, with similar overall properties, including similar critical exponents. This indicates that a critical role in originating the Barkhausen noise is played by disorder, yet it does not seem to be crucial whether disorder is realized through quenched random fields (RFIM) or though randomly varying coupling constants (RBIM).

\end{document}